\documentclass[10pt,floatfix,prd,superscriptaddress,nofootinbib,amsmath,amssymb,aps,twocolumn]{revtex4-1}
\usepackage{graphicx}
\usepackage{xcolor}
\usepackage{bm}
\usepackage[hypertexnames=false,hidelinks]{hyperref}
\usepackage{xspace}
\usepackage{tikz}
\usetikzlibrary{arrows.meta,positioning,shapes.geometric}

\allowdisplaybreaks[4]

\begin{document}

\title{AI-Driven Discovery of Information-Efficient Collider Observables for Interference Measurements}
 \author{Jiahui Lin}

  \affiliation{School of Physics and Astronomy, Beijing Normal University, and Key Laboratory of Multiscale Spin Physics (Beijing Normal University), Ministry of Education, Beijing 100875, China}

   \author{Yandong Liu}
    \email{ydliu@bnu.edu.cn} 

    \affiliation{School of Physics and Astronomy, Beijing Normal University, and Key Laboratory of Multiscale Spin Physics (Beijing Normal University), Ministry of Education, Beijing 100875, China}

\begin{abstract}

Optimal observables provide statistically powerful probes of small deformations from a reference theory, but in realistic collider measurements they are rarely available in compact analytic form. We show that interpretable event-level observables can be discovered by AI-driven symbolic evolution using score information from matrix-element reweighting as the statistical target. Focusing on the CP-sensitive interaction $HZ_{\mu\nu}\tilde Z^{\mu\nu}$, we study two complementary realizations of the same coupling structure: associated production $e^+e^-\to Z(\to \mu^-\mu^+)H$ and the decay channel $pp\to H\to ZZ^*\to e^-e^+\mu^-\mu^+$. The learned observables retain substantially more local Fisher information than standard angular baselines while remaining compact analytic functions. In both cases, the discovered expressions recover characteristic helicity-interference harmonics. In associated production these harmonics are supplemented by laboratory-frame asymmetry mappings, while in four-lepton decay the robust component is the angular kernel, with the mass-ratio factor serving as a bounded representative prefactor. These results recast optimal-observable design as a symbolic discovery problem and provide a transparent route to information-efficient, interpretable probes of collider interference.

\end{abstract}

\maketitle

\section{Introduction}

Future collider programs are entering a precision regime in which the central challenge is to discriminate among subtle deformations of the Standard Model using high-dimensional event information. In this setting, sensitivity is often controlled by interference, and the choice of observable becomes a primary ingredient of the measurement strategy. In principle, the optimal event-level probe is the score function, which saturates the Fisher-information bound~\cite{Davier:1993jp,Diehl:1994fz,Brehmer:2019xox}. In practice, however, the score is rarely available in a compact analytic form that is transparent, portable, and experimentally robust.

From this information-theoretic viewpoint, the full event-level distribution contains the maximal local Fisher information about the parameter of interest. Any experimentally useful observable is a compression of the event kinematics and generally retains only part of that information. A higher Fisher-information efficiency therefore means that less parameter-relevant information is lost in the compression, leading to stronger statistical sensitivity at fixed data size.

This tension has motivated two complementary lines of development. Analytic observables are physically interpretable and often experimentally stable, but are usually constructed by hand and may miss important correlations across phase space. Modern machine-learning approaches can instead approach optimal statistical performance by exploiting high-dimensional features, but typically do so through black-box architectures whose learned structure is difficult to interpret physically~\cite{Guest_2018,Cid_Vidal_2021,Larkoski:2017jix,Radovic:2018dip,Collins:2019jip,Ren:2017ymm,Abdughani:2020xfo,Ren:2021prq,Lu:2023gjk,Brehmer:2019xox,Cruz:2024grk,Barrue:2023ysk,Butter:2022rso}. This motivates the central goal of this work: to construct compact, interpretable observables that retain parameter-relevant score information while remaining analytic and experimentally portable.

Symbolic regression offers a natural intermediate route, aiming to learn analytic surrogates for performant observables while preserving interpretability~\cite{Butter:2021rvz,Bahl:2025jtk}. More recently, AI-driven evolutionary search, exemplified by frameworks such as FunSearch~\cite{funsearch} and AlphaEvolve~\cite{novikov2025alphaevolvecodingagentscientific}, has demonstrated that structured mathematical objects can be discovered efficiently in program space rather than by brute-force exploration of a fixed functional basis~\cite{Song:2025pwy,Cao:2025shc}. The collider-observable problem is particularly well suited to this perspective: the target is not an arbitrary predictor, but a compact analytic function constrained by symmetry, kinematics, and statistical optimality.

In this Letter, we show that information-efficient and interpretable collider observables can be discovered from event-level kinematics by AI-driven symbolic evolution, with matrix-element reweighting~\cite{Artoisenet:2010cn} providing the local score target used for optimization, in line with the broader program of simulation-based inference for particle physics~\cite{Cranmer:2020zzo}. We study a CP-sensitive deformation of the Higgs--$Z$ interaction,
\begin{equation}
\mathcal{L} \supset \mathcal{L}_{\rm SM} + \frac{\kappa}{\Lambda} H Z_{\mu\nu}\tilde Z^{\mu\nu},
\label{eq:lagrangian}
\end{equation}
and consider two distinct collider realizations of the same interaction: associated production $e^+e^-\to Z(\to \mu^-\mu^+)H$ and the decay channel $pp\to H\to ZZ^*\to e^-e^+\mu^-\mu^+$, both of which are established probes of Higgs tensor structure and CP properties~\cite{Gounaris:1996zh,Mahlon:1998zh,Beneke:2014sba,Craig:2015wwr,Sha:2022lbb,Gao:2010zz,Bolognesi:2012zz,Anderson:2014hqa,Godbole:2007cn,Chen:2013ejz,Chen:2014gka}. In both cases, the symbolic search identifies compact analytic observables that track the score substantially better than conventional angular baselines. The learned expressions contain CP-sensitive interference kernels supplemented by process-dependent signed mappings or smooth prefactors that improve or stabilize the one-dimensional projection onto the score. This shows that optimal-observable design can be reformulated as a symbolic discovery problem, and that the resulting expressions reveal the underlying interference geometry in analytic form.

\section{Setup and strategy}

Local sensitivity to a parameter $\kappa$ is captured by the Fisher information $I(\kappa)=\mathbb{E}\!\left[\left(\partial_\kappa\ln p(x|\kappa)\right)^2\right]$, which sets the Cram\'er--Rao lower bound $\mathrm{Var}(\hat\kappa)\geq 1/I(\kappa)$ on the variance of any unbiased estimator. The score $s(x)=\partial_\kappa\ln p(x|\kappa)|_{\kappa=0}$ saturates this bound and is therefore the optimal event-level observable in the small-deformation limit. Classical ``optimal-observable'' constructions in collider phenomenology are projections onto this direction~\cite{Atwood:1991ka,Davier:1993jp,Diehl:1994fz,Babich:1999kp}. For a one-dimensional compression $y=f(x)$, the retained Fisher information is $I[f]=\mathbb{E}\!\left[\left(\mathbb{E}[s|y]\right)^2\right]$, and the Fisher efficiency
\begin{equation}
\epsilon[f]=\frac{I[f]}{I_{\rm full}}\in[0,1]
\end{equation}
quantifies how much of the per-event information survives the compression.
Because the observables considered here are compact one-dimensional
compressions, $\epsilon$ need not be close to unity; we compare it to standard
interpretable baselines.

For event-level measurements of $\kappa$ in Eq.~(\ref{eq:lagrangian}), the differential event weight may be expanded around the Standard Model point as
\begin{equation}
w(x|\kappa)=c_0(x)+\kappa\,c_1(x)+\kappa^2 c_2(x)+\mathcal{O}(\kappa^3),
\end{equation}
where $c_0$ denotes the CP-even contribution, $c_1$ the interference term linear in $\kappa$, and $c_2$ the quadratic contribution. The locally optimal event-level statistic at the Standard Model point is then
\begin{equation}
t(x)=\frac{c_1(x)}{c_0(x)},
\end{equation}
which coincides with the score of the normalized distribution up to an $x$-independent shift~\cite{Brehmer:2019xox,Bhattacharya:2025jhs}.

\begin{figure*}[t]
    \centering
    \begin{tikzpicture}[
        node distance=5mm,
        >=Latex,
        box/.style={
            rectangle,
            rounded corners=3pt,
            draw=black,
            semithick,
            align=center,
            text width=0.21\textwidth,
            minimum height=1.7cm,
            inner sep=3.5pt
        },
        inputbox/.style={box, fill=blue!10},
        targetbox/.style={box, fill=cyan!10},
        searchbox/.style={box, fill=green!10},
        outputbox/.style={box, fill=orange!12},
        flow/.style={->, semithick}
    ]
        \node[inputbox] (input) {\textbf{Event Data}\\
        Kinematics $x$ near a reference theory};

        \node[targetbox, right=of input] (score) {\textbf{Sensitivity Target}\\
        Local score $t(x)=c_1/c_0$ from MG5 reweighting};

        \node[searchbox, right=of score] (search) {\textbf{AI-Driven Symbolic Search}\\
        Fisher-efficiency fitness, LLM-guided evolution};

        \node[outputbox, right=of search] (output) {\textbf{Interpretable Observable}\\
        kernel $\times$ frame mapping $\times$ kinematic weight};

        \draw[flow] (input) -- (score);
        \draw[flow] (score) -- (search);
        \draw[flow] (search) -- (output);
    \end{tikzpicture}
  \caption{Schematic overview of the AI-driven symbolic-observable construction. A local score target derived from matrix-element reweighting drives an LLM-guided evolutionary search in a physics-motivated function space, and the final observable is selected by its Fisher-information efficiency $\epsilon[f]$.}
    \label{fig:workflow}
\end{figure*}
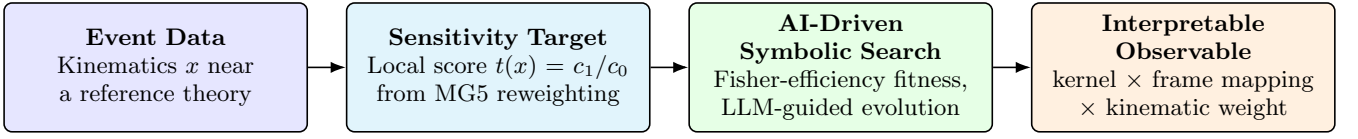

Rather than deriving $t(x)$ analytically and then seeking simplified approximations, we directly search for compact analytic functions $f(x)$ that maximize $\epsilon[f]$ (Fig.~\ref{fig:workflow}). The score is estimated numerically by matrix-element reweighting, while the symbolic search is restricted to a physics-motivated function space built from measured kinematic variables and simple analytic operations. This strategy does not assume analytic amplitude input in the construction of candidate observables, but uses the score only as the target statistical direction against which the symbolic expressions are evaluated.

We study two complementary realizations of the same interaction: $e^+e^-\to Z(\to \mu^-\mu^+)H$ at a lepton collider, and $pp\to H\to ZZ^*\to e^-e^+\mu^-\mu^+$ at a hadron collider. These benchmarks allow us to compare the learned observables across distinct production and decay kinematics. All numerical results reported below are evaluated on truth-level matrix-element-reweighted samples; parton-shower, detector, and continuum-background effects are left to follow-up work.

For each channel, we compare the learned symbolic observables with standard angular baselines using $\epsilon[f]$ of the corresponding one-dimensional compression. We then analyze the resulting expressions to identify the physical ingredients that are stable across collider realizations.

\begin{table*}[t]
\caption{Compact summary of the symbolic observables.  The quoted efficiencies are the Fisher information retained after one-dimensional compression, normalized to the full local score information in the corresponding truth-level benchmark.}
\begin{ruledtabular}
\begin{tabular}{p{0.19\textwidth}p{0.22\textwidth}p{0.14\textwidth}p{0.35\textwidth}}
Channel & Baseline efficiency & Symbolic efficiency & Stable symbolic structure \\
\hline
$e^+e^-\to ZH$ & $0.059$ for $\sin(2\phi^*)$ & $0.102$ &
interference kernels $\times$ charge-ordered lab-frame mappings and smooth $p_T^Z$ prefactor \\
$H\to ZZ^*\to4\ell$ & $2.3\times10^{-4}$ for $\sin(\phi_1^*+\phi_2^*)$ & $1.9\times10^{-2}$ &
CP-sensitive angular interference kernel; smooth bounded representative prefactor
\end{tabular}
\end{ruledtabular}
\end{table*}

\section{Emergent structure of the discovered observables}

\subsection{Associated production \texorpdfstring{$e^+e^-\to Z(\to \mu^-\mu^+)H$}{e+e- -> Z(mu-mu+)H}}

\begin{figure}[!tbp]
    \centering
    \includegraphics[width=\linewidth]{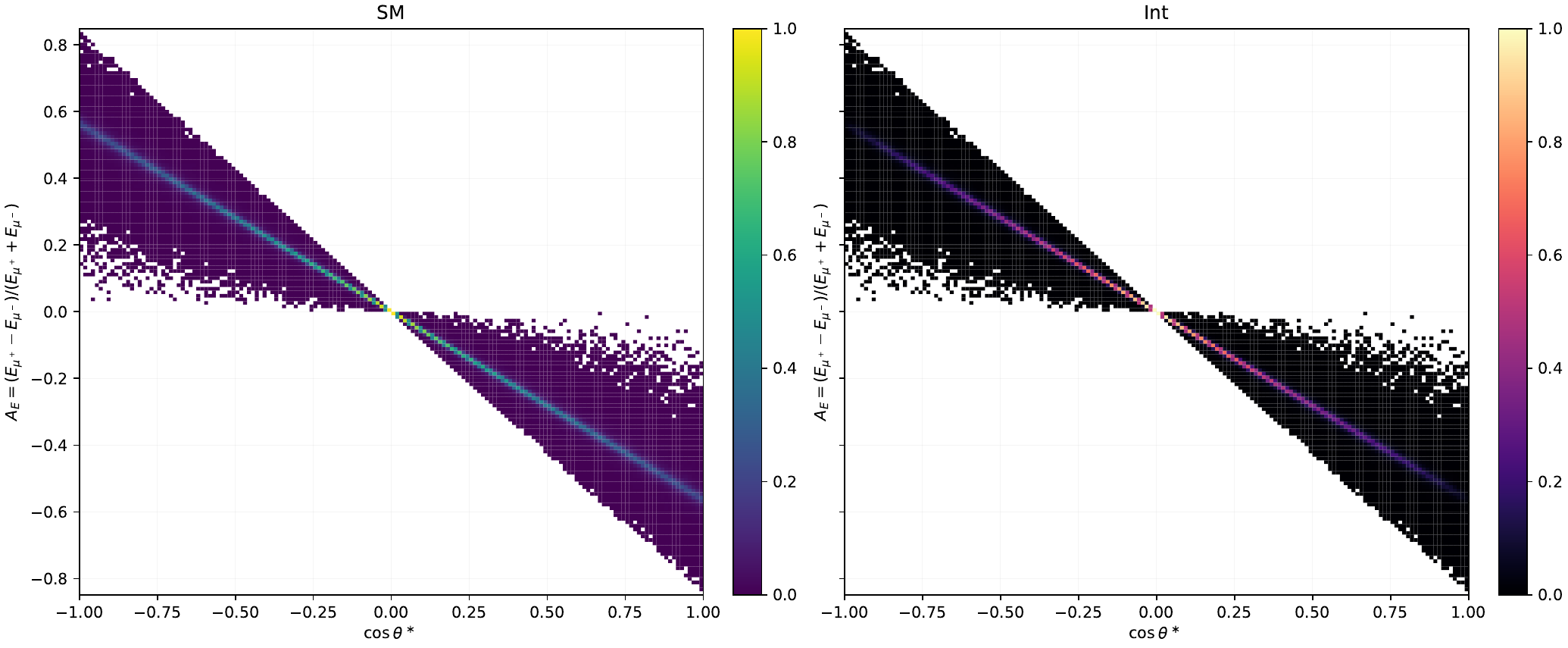}
  \caption{Laboratory-frame energy asymmetry
$A_E=(E_{\mu^+}-E_{\mu^-})/(E_{\mu^+}+E_{\mu^-})$ versus $\cos\theta^\ast$ in $e^+e^-\to Z(\to\mu^-\mu^+)H$ at $\sqrt{s}=250$ GeV. Left: Standard Model event density. Right: the same plane weighted by $|c_1|$. The narrow band in both panels shows that $A_E$ is an effective proxy for $\cos\theta^\ast$.
}
    \label{fig:AE_vs_costheta_star}
\end{figure}

\begin{figure}[!tbp]
    \centering
    \includegraphics[width=\linewidth]{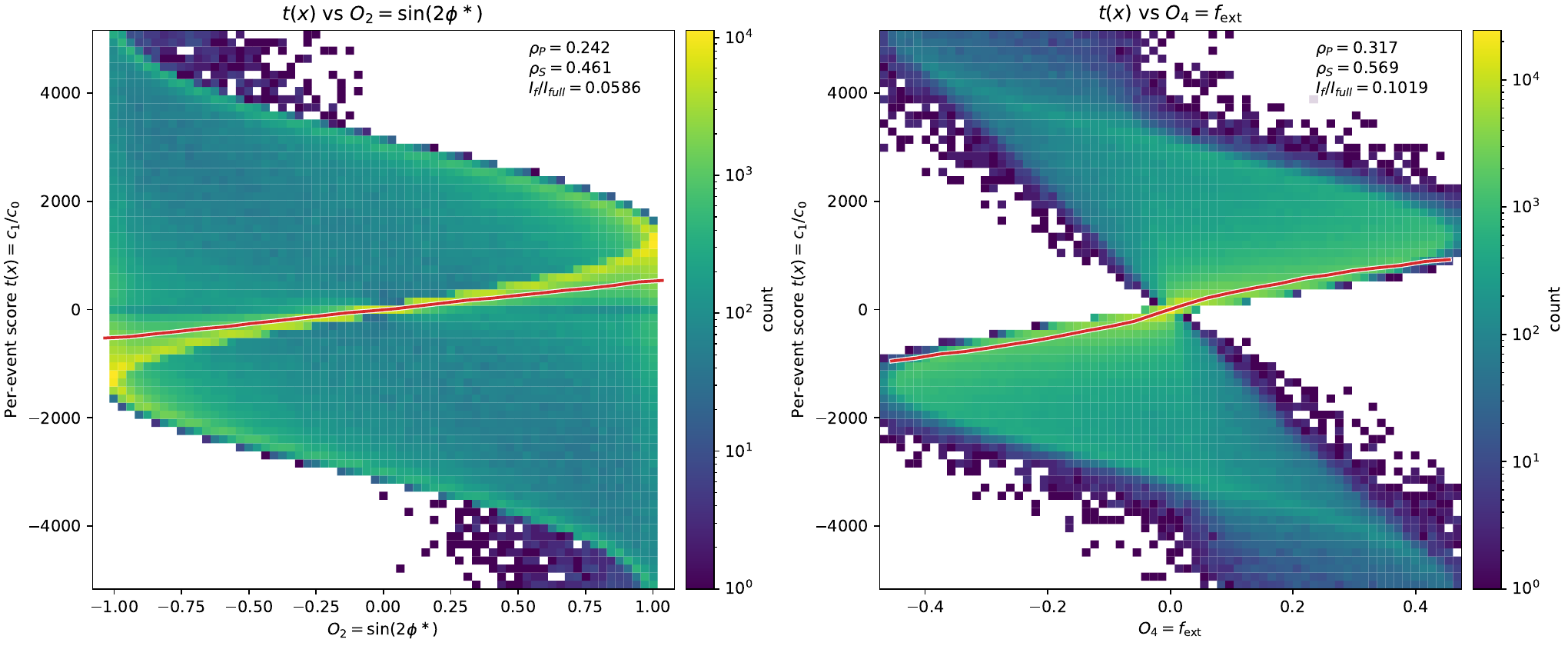}
  \caption{Per-event score $t(x)=c_1/c_0$ versus two observables in $e^+e^-\to Z(\to\mu^-\mu^+)H$ at $\sqrt{s}=250$ GeV. Left: $O_2=\sin(2\phi^\ast)$ with $\epsilon\simeq 0.059$. Right: $O_4=f_{\rm ext}$ with $\epsilon\simeq 0.102$. The learned observable follows the score more monotonically and therefore provides a more faithful one-dimensional compression.
}
    \label{fig:scorevsobservables}
\end{figure}

\begin{figure*}[t]
    \centering
    \includegraphics[width=0.95\textwidth]{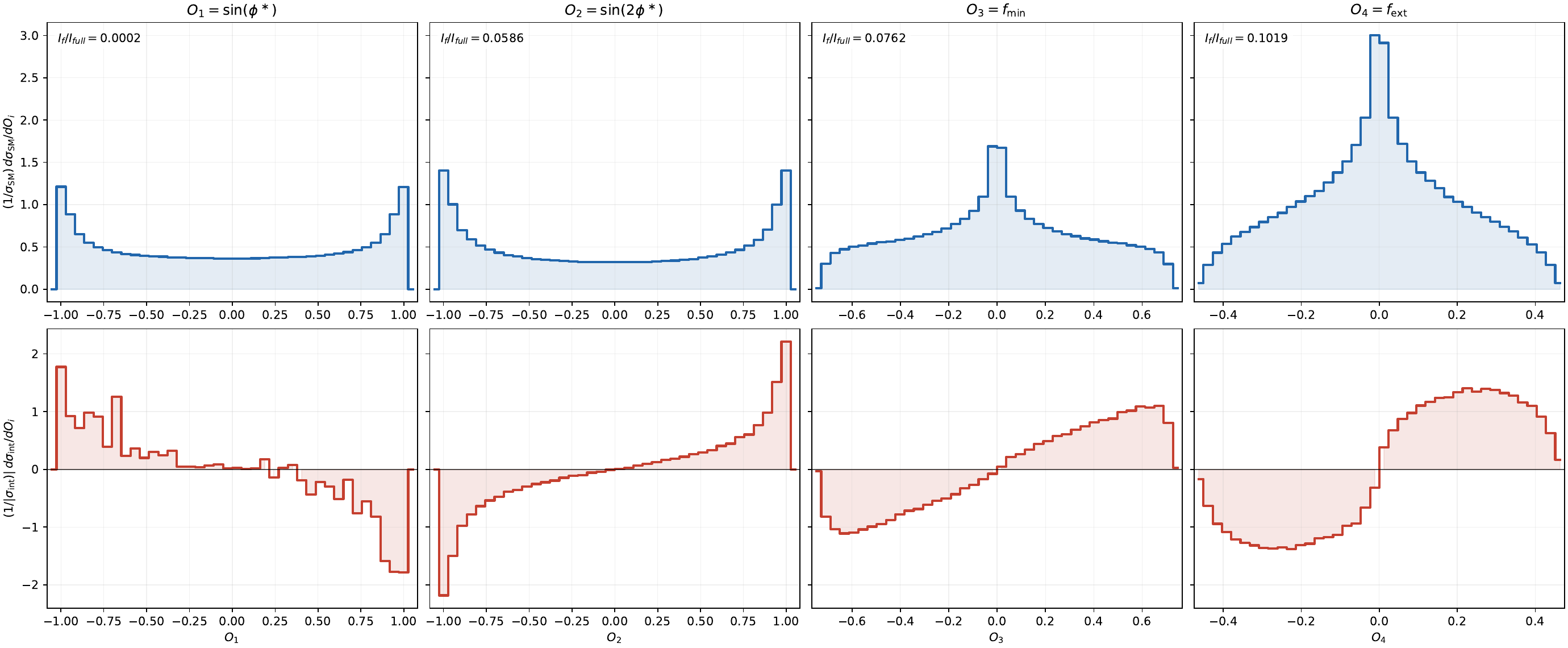}
  \caption{Standard Model and signed interference distributions for
$O_1=\sin\phi^\ast$,
$O_2=\sin(2\phi^\ast)$,
$O_3=f_{\rm min}$, and
$O_4=f_{\rm ext}$ in $e^+e^-\to Z(\to\mu^-\mu^+)H$ at $\sqrt{s}=250$ GeV. From $O_1$ to $O_4$, opposite-sign interference regions become progressively better separated. The corresponding efficiencies are $\epsilon\simeq 2.0\times10^{-4}$, $0.059$, $0.076$, and $0.102$.
}
 \label{fig:distributions}
\end{figure*}

We first consider the minimal angular description
\begin{equation}
x=(\cos\theta_Z,\cos\theta^*,\phi^*),
\end{equation}
where $\theta_Z$ is the production angle of the reconstructed $Z$ boson, and $(\theta^*,\phi^*)$ are the decay angles of the negatively charged muon. The precise helicity-frame axes are defined in Appendix~\ref{sec:kinematic_conventions}. Within this basis, the symbolic search identifies an observable of the form
\begin{equation}
f_{\rm min}(x)\sim
\frac{\sin^2\theta_Z\sin^2\theta^*\sin(2\phi^*)}
{\sqrt{0.5+(\sin^2\theta_Z\sin^2\theta^*)^2(1.5+\cos2\phi^*)}}.
\end{equation}
This minimal observable already reaches a Fisher-information efficiency of $\epsilon\simeq 0.076$, clearly above the best simple angular benchmark $\sin(2\phi^*)$ with $\epsilon\simeq 0.059$.
Its leading numerator structure,
\begin{equation}
\sin^2\theta_Z\sin^2\theta^*\sin(2\phi^*),
\end{equation}
reproduces the characteristic CP-sensitive harmonic associated with transverse--transverse helicity interference~\cite{Gounaris:1996zh,Mahlon:1998zh,Beneke:2014sba,Craig:2015wwr}. The result is nontrivial: this angular pattern is not imposed in the symbolic search, but emerges from direct optimization of the Fisher objective.

The structure becomes more revealing when laboratory-frame information is included. In the extended feature representation, with $x'$ denoting the angular variables augmented by the laboratory-frame quantities defined in Appendix~\ref{sec:kinematic_conventions}, the best symbolic observable takes the schematic form
\begin{align}\label{eq:fext}
f_{\mathrm{ext}}(x')
&=
\Big[
0.706 \, \sin(2\phi^*) \sin^2\theta^* \sin\theta_Z
\nonumber\\
&\quad
+ \sin\phi^* \sin\theta^* \cos\theta_Z \left(2A_E + A_{p_T}\right)
\Big] \nonumber \\
&\quad \times \frac{(p_T^{Z})^2}{(p_T^{Z })^2 + p_0^2} \, ,
\end{align}
where $A_E$, $A_{p_T}$, $p_T^Z$, and $p_0=46~\mathrm{GeV}$ are defined in Appendix~\ref{sec:kinematic_conventions}. The numerical constants are selected by the search.
Compared with conventional angular observables such as $\sin(2\phi^*)$, this expression tracks the score much more closely, and its structure admits a helicity-interference interpretation.

First, the angular component retains a CP-sensitive interference harmonic. Second, the laboratory-frame asymmetry $A_E$ supplies an experimentally accessible proxy for the decay polar factor. At fixed $\sqrt{s}$, the two-body kinematics imply $A_E=(E_{\mu^+}-E_{\mu^-})/(E_{\mu^+}+E_{\mu^-})=-\beta_Z\cos\theta^*$ up to lepton-mass corrections. In Eq.~(\ref{eq:fext}), this proxy appears in the product $\sin\phi^*\sin\theta^*\cos\theta_Z A_E\simeq-(\beta_Z/2)\cos\theta_Z\sin(2\theta^*)\sin\phi^*$. In this product, $A_E$ supplies the decay-side $\cos\theta^*$ sign, while $\cos\theta_Z$ supplies the forward--backward production sign, yielding the signed pattern $\cos\theta_Z\sin\theta^*\cos\theta^*\sin\phi^*$. This is one of the angular factors entering the longitudinal--transverse helicity interference, as detailed in Appendix~\ref{sec:helicity_structure}. The $A_E$--$\cos\theta^*$ mapping is shown explicitly in Fig.~\ref{fig:AE_vs_costheta_star}, where both the Standard Model density and the $|c_1|$-weighted density follow a narrow approximately linear band.

The transverse-momentum asymmetry $A_{p_T}$ is the corresponding charge-ordered laboratory-frame lepton-$p_T$ asymmetry and appears within the same laboratory-frame mapping that multiplies the CP-sensitive angular structure. The factor $(p_T^Z)^2/[(p_T^Z)^2+p_0^2]$ acts as a smooth saturating prefactor that enhances central production configurations and stabilizes the observable as $p_T^Z\to 0$.

Altogether, the extended observable reaches $\epsilon\simeq 0.102$, improving the information efficiency by about $74\%$ relative to the best simple angular baseline. Figure~\ref{fig:scorevsobservables} shows this improvement directly: compared with $O_2=\sin(2\phi^*)$, the learned observable $f_{\rm ext}$ follows the score much more monotonically. Figure~\ref{fig:distributions} shows the corresponding progression within this channel from $O_1$ to $O_4$, where opposite-sign interference regions become increasingly well separated. The discovered observable therefore combines CP-sensitive angular harmonics, a signed production--decay angular factor, and a smooth kinematic prefactor in a single compact analytic expression.

\subsection{Higgs decay \texorpdfstring{$pp\to H\to ZZ^*\to e^-e^+\mu^-\mu^+$}{pp -> H -> ZZ* -> e-e+mu-mu+}}

We next consider the four-lepton decay channel, which probes the same interaction in a kinematically distinct realization. Here the symbolic search identifies an observable of the form
\begin{equation}
O_{4\ell}=r_Z(1-r_Z)\sin(2\theta_1^*)\sin(2\theta_2^*)\sin(\phi_1^*+\phi_2^*),
\end{equation}
where $r_Z\equiv m_{Z_2}/(m_{Z_1}+m_{Z_2})$, with the fixed-flavor convention $Z_1\equiv e^-e^+$ and $Z_2\equiv\mu^-\mu^+$. The angle and transverse-axis conventions for $(\theta_i^*,\phi_i^*)$ are defined in Appendix~\ref{sec:kinematic_conventions}.
Again, the angular dependence reconstructs the expected CP-sensitive interference harmonic, now appropriate to the $H\to ZZ^*\to 4\ell$ decay geometry~\cite{Choi:2002jk,Gao:2010zz,Bolognesi:2012zz,Anderson:2014hqa,Belyaev:2015xwa,Chen:2014gka,Avery:2012um}. With the angles $(\theta_i^*,\phi_i^*)$ defined in each $Z_i$ rest frame, where the two $Z$ bosons are back-to-back, the azimuthal sum $\phi_1^*+\phi_2^*$ is precisely the dihedral angle between the two decay planes, and is the standard CP-odd azimuthal baseline of MELA-type four-lepton analyses~\cite{Gao:2010zz,Anderson:2014hqa}. The polar factor $\sin(2\theta_1^*)\sin(2\theta_2^*)$ isolates the relevant longitudinal--transverse interference.

The accompanying invariant-mass factor $r_Z(1-r_Z)$ is a smooth, bounded representative prefactor selected by the symbolic search. Direct ablations indicate that the Fisher efficiency is dominated by the angular kernel, so the physical interpretation below focuses on the CP-sensitive angular structure.

\begin{figure}[!tbp]
    \centering
    \includegraphics[width=\linewidth]{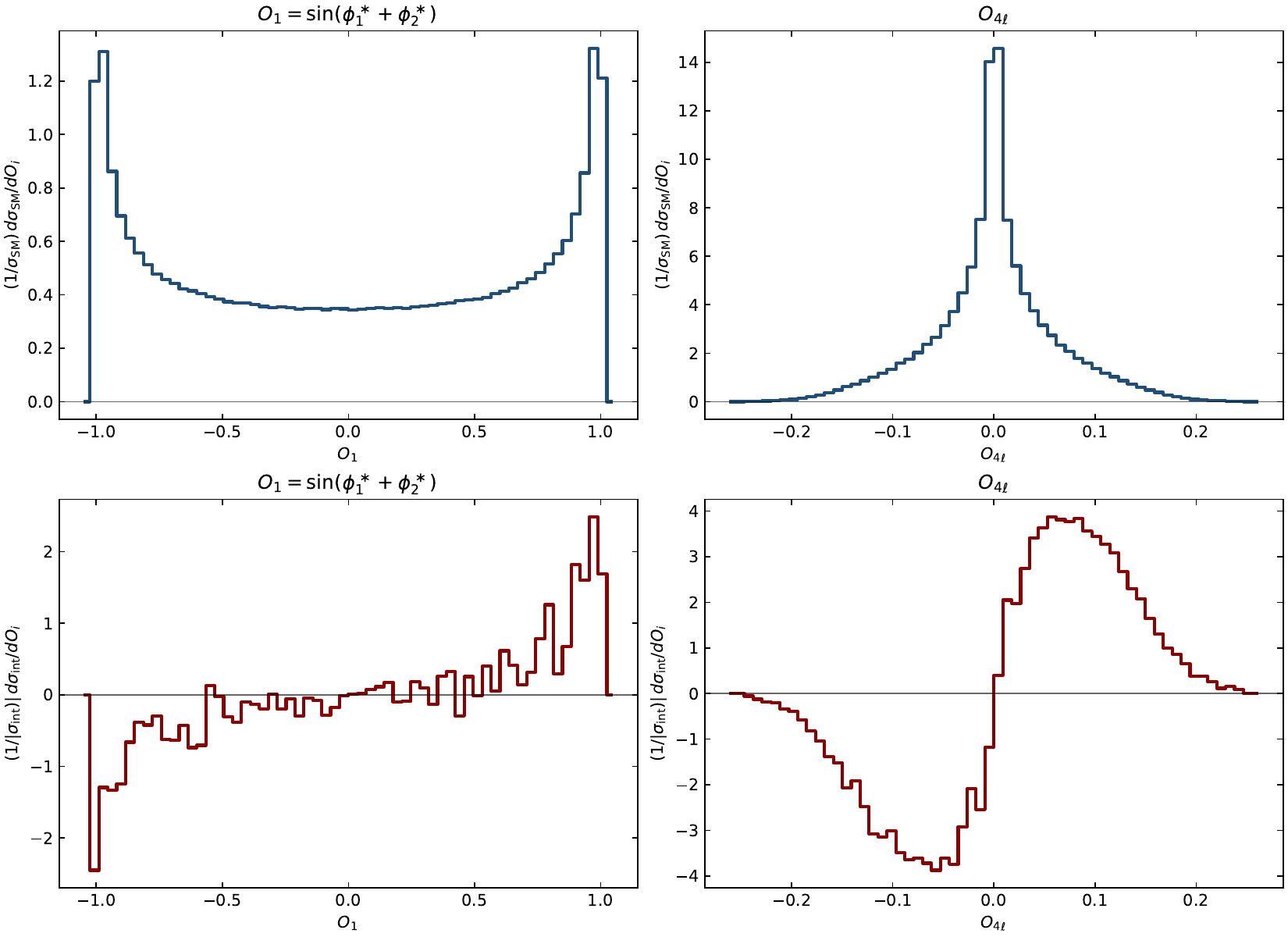}
    \caption{Standard Model and signed interference distributions for $O_1=\sin(\phi_1^\ast+\phi_2^\ast)$ and $O_{4\ell}$ in $pp\to H \to ZZ^\ast\to e^-e^+\mu^-\mu^+$. The symbolic observable separates opposite-sign interference regions much more clearly than the angular baseline. The Fisher-information efficiency improves from $\epsilon\simeq 2.3\times10^{-4}$ to $\epsilon\simeq 1.9\times10^{-2}$.}
    \label{fig:ppdistributions}
\end{figure}

In this truth-level signal-only benchmark, the best simple angular baseline $O_1=\sin(\phi_1^*+\phi_2^*)$ carries only a tiny fraction of the available information, with $\epsilon\simeq 2.3\times 10^{-4}$, whereas the symbolic observable reaches $\epsilon\simeq 1.9\times 10^{-2}$. Figure~\ref{fig:ppdistributions} shows this improvement directly: the learned observable reorganizes the events so that opposite-sign interference regions occupy much more distinct parts of the observable axis than for the pure angular baseline. Interpreting the learned observable as a signed asymmetry variable, we further find that a moderate cut around $|O_{4\ell}|\simeq 0.05$ improves an asymmetry-based $S_{\rm asym}/\sqrt{D_{\rm SM}}$ measure by about $30\%$, where $D_{\rm SM}$ is the Standard Model contribution within the same signal process and no continuum backgrounds are included; the definition and scan are summarized in the Appendix.

\subsection{Interference kernels and process-dependent mappings}

The two collider realizations exhibit an interference-kernel structure that can be written schematically as
\begin{align}
f(x)&\sim
\bigl({\rm interference\ kernel}\bigr) \nonumber\\
&\quad\times
\bigl({\rm process\mbox{-}dependent\ mapping\ or\ prefactor}\bigr).
\end{align}
Within these two truth-level benchmarks, the recovered observables can be interpreted as analytic combinations of distinct physical ingredients: the helicity-interference harmonic responsible for CP sensitivity, and process-dependent mappings or prefactors that expose signed angular structure, improve the projection, or provide bounded representative forms.

This pattern also clarifies the role of symbolic search in collider analysis. The results support the interpretation that the search identifies analytic compressions of event information that align with the score because they capture the interference geometry in experimentally accessible variables. Figures~\ref{fig:distributions} and \ref{fig:ppdistributions} illustrate this behavior: as the observables improve, the positive and negative interference regions become progressively better separated along the one-dimensional observable axis.

\section{Conclusion}

We have shown that information-efficient and interpretable collider observables can be discovered from event-level kinematics by AI-driven symbolic evolution with a Fisher-information objective. For the CP-sensitive interaction $HZ_{\mu\nu}\tilde Z^{\mu\nu}$, the learned observables in both $e^+e^-\to Z(\to \mu^-\mu^+)H$ and $pp\to H\to ZZ^*\to e^-e^+\mu^-\mu^+$ track the score substantially better than conventional angular baselines while remaining compact and analytic. In associated production, the Fisher-information efficiency improves from $\epsilon\simeq 0.059$ for the best simple angular baseline to $\epsilon\simeq 0.102$ for the best symbolic observable.

The central result is that the learned observables contain interference kernels supplemented by process-dependent signed mappings or smooth prefactors. The optimal-observable problem can therefore be reformulated as a symbolic discovery problem, in which interpretable analytic expressions emerge directly from the geometry of interference in phase space.

This perspective provides a transparent bridge between traditional analytic observable design and modern data-driven optimization. More broadly, it suggests that the structure of statistically efficient measurements can itself be discovered and interpreted as a physical object, opening a systematic route toward information-efficient observables in precision collider studies. The results reported here are truth-level local Fisher benchmarks; incorporating parton shower, detector response, and continuum backgrounds, and extending the search to likelihood-oriented objectives, are natural next steps.

\begin{acknowledgments}
 The work of Y.L. is partly supported by the National Science Foundation of China under Grant Nos. 12075257 and 12175016, the National Key R$\&$D Program of China under Grant No. 2023YFA1607104, and Fundamental Research Funds for the Central Universities, Beijing
Normal University.

\end{acknowledgments}

\bibliographystyle{apsrev}
\bibliography{ref}

\clearpage
\appendix
\onecolumngrid
\section{Score Estimation and Fisher-Information Projection}
\label{sec:score_estimation}

This appendix is organized around the points needed to reproduce and interpret
the main-text results.  Sec.~\ref{sec:score_estimation} defines the score and
projected Fisher-information estimator.  Sec.~\ref{sec:samples_observables}
collects the event samples, benchmark efficiencies, and the explicit
observables quoted in the main text.  Sec.~\ref{sec:kinematic_conventions}
fixes the kinematic conventions.  Sec.~\ref{sec:helicity_structure} explains
the helicity-interference angular factors used to interpret the $ZH$
observable.  Sec.~\ref{sec:search_robustness} documents the symbolic search and
robustness checks.  Sec.~\ref{sec:asymmetry_selection} gives the four-lepton
asymmetry-selection diagnostic quoted in the main text.

For a single deformation parameter $\kappa$, the event weight is expanded around the reference point as
\begin{align}
w(x|\kappa)=c_0(x)+\kappa c_1(x)+\kappa^2 c_2(x)+\mathcal{O}(\kappa^3).
\end{align}
Here $x$ denotes the full event-level kinematic configuration and
$\sigma(\kappa)=\int dx\,w(x|\kappa)$, so that
$p(x|\kappa)=w(x|\kappa)/\sigma(\kappa)$.
The score of the normalized distribution is
\begin{align}
s(x)=\left.\frac{\partial}{\partial\kappa}\ln p(x|\kappa)\right|_{\kappa=0}
=\frac{c_1(x)}{c_0(x)}-\frac{\sigma'(0)}{\sigma(0)} .
\end{align}
Thus the uncentered ratio
\begin{align}
t(x)=\frac{c_1(x)}{c_0(x)}
\end{align}
differs from the exact score only by an $x$-independent shift and is sufficient for local one-dimensional shape compression.

In the symmetric reweighting setup used for the CP-odd studies, event weights are available at $\kappa=0,\pm\delta$:
\begin{align}
w^0(x)&=c_0(x),\\
w^+(x)&=c_0(x)+\delta c_1(x)+\delta^2 c_2(x),\\
w^-(x)&=c_0(x)-\delta c_1(x)+\delta^2 c_2(x).
\end{align}
The coefficient estimator is therefore
\begin{align}
c_1(x)\simeq \frac{w^+(x)-w^-(x)}{2\delta},
\qquad
c_2(x)\simeq \frac{w^+(x)+w^-(x)-2w^0(x)}{2\delta^2}.
\end{align}
If weights are instead available at three nonsymmetric coupling values, the coefficients should be obtained by solving the corresponding three-point linear system.

For a projected observable $y=f(x)$, the one-dimensional information estimator
used for the quoted benchmarks is
\begin{align}
I[f]=\mathbb{E}\!\left[\left(\mathbb{E}[t|y]\right)^2\right],
\end{align}
with equal-population bins in $y$ used for the numerical evaluation.
The reported Fisher efficiency is
\begin{align}
\epsilon[f]=\frac{I[f]}{I_{\rm full}},
\qquad
I_{\rm full}=\langle t^2\rangle_0 .
\end{align}
For the exact Fisher information of the normalized shape, $t$ should be
replaced by $\tilde t=t-\langle t\rangle_0$.  In the two CP-odd benchmarks
used here, $\langle t\rangle_0^2/\langle t^2\rangle_0$ is below
$1.5\times10^{-5}$, so this distinction is numerically invisible at the
precision quoted in the tables.

The same estimator can be written in the weighted-histogram form used in the
analysis workflow.  If the compressed observable is divided into bins $B_k$, define
\begin{align}
\sigma_k^{(0)}=\sum_{i\in B_k}w_i^{(0)},\qquad
\sigma_k^{(1)}=\sum_{i\in B_k}w_i^{(1)},\qquad
\sigma_{\rm tot}^{(0)}=\sum_i w_i^{(0)} .
\end{align}
Here $w_i^{(0)}\equiv c_0(x_i)$ is the reference contribution and
$w_i^{(1)}\equiv c_1(x_i)$ is the linear interference coefficient for event
$i$.
Then
\begin{align}
I[f]\simeq \frac{1}{\sigma_{\rm tot}^{(0)}}
\sum_k \frac{\left(\sigma_k^{(1)}\right)^2}{\sigma_k^{(0)}} .
\end{align}
For approximately unweighted Standard-Model samples this reduces to
\begin{align}
I[f]\simeq \frac{1}{N}\sum_k
\frac{\left(\sum_{i\in B_k}t_i\right)^2}{n_k},
\end{align}
where $n_k$ is the number of events in bin $B_k$.  Equal-population binning is
used unless otherwise stated, because it avoids empty bins for sharply peaked
or asymmetric symbolic observables.

\section{Analysis Samples and Benchmark Observables}
\label{sec:samples_observables}

The CP-odd studies use truth-level matrix-element reweighted samples.  The
local score is used only as the optimization target and for benchmarking the
one-dimensional compression; it is not inserted into the candidate observable
itself.  The numerical samples used for the quoted efficiencies are summarized
in Table~\ref{tab:samples}.

\begin{table}[ht]
\caption{Truth-level samples used for the local Fisher-information benchmarks.}
\label{tab:samples}
\begin{center}
\begin{tabular}{lcccc}
\hline
Channel & Events & Reweighting step & $I_{\rm full}$ & Nominal bins \\
\hline
$e^+e^-\to Z(\to\mu^-\mu^+)H$ at $\sqrt{s}=250~{\rm GeV}$ &
$1.0\times10^5$ & $2.5\times10^{-4}$ & $2.38\times10^6$ & $30$ \\
$pp\to H\to ZZ^*\to e^-e^+\mu^-\mu^+$ &
$1.0\times10^6$ & $2.0\times10^{-3}$ & $5.98\times10^4$ & $30$ \\
\hline
\end{tabular}
\end{center}
\end{table}

Table~\ref{tab:benchmark_details} gives the benchmark efficiencies
corresponding to the compact summary in the main text.  The explicit symbolic
expressions quoted in the main text are collected below the table; their
kinematic variables are defined in Sec.~\ref{sec:kinematic_conventions}.

\begin{table}[ht]
\caption{One-dimensional observables and Fisher-information efficiencies used
in the main text.}
\label{tab:benchmark_details}
\begin{center}
\begin{tabular}{lcc}
\hline
Channel & Observable & $\epsilon$ \\
\hline
$ZH$ & $\sin(2\phi^*)$ baseline & $0.059$ \\
$ZH$ & minimal symbolic $f_{\rm min}$ & $0.076$ \\
$ZH$ & extended symbolic $f_{\rm ext}$ & $0.102$ \\
$4\ell$ & $\sin(\phi_1^*+\phi_2^*)$ baseline & $2.3\times10^{-4}$ \\
$4\ell$ & symbolic $O_{4\ell}$ & $1.9\times10^{-2}$ \\
\hline
\end{tabular}
\end{center}
\end{table}

The two symbolic $ZH$ representatives are
\begin{align}
f_{\rm min}
&\sim
\frac{\sin^2\theta_Z\sin^2\theta^*\sin(2\phi^*)}
{\sqrt{0.5+\left(\sin^2\theta_Z\sin^2\theta^*\right)^2
\,\left(1.5+\cos2\phi^*\right)}},\\
f_{\rm ext}
&=
\left[
0.706\sin(2\phi^*)\sin^2\theta^*\sin\theta_Z
+\sin\phi^*\sin\theta^*\cos\theta_Z(2A_E+A_{p_T})
\right]
\frac{(p_T^Z)^2}{(p_T^Z)^2+p_0^2}.
\end{align}
For the four-lepton channel,
\begin{align}
O_{4\ell}=
r_Z(1-r_Z)\sin(2\theta_1^*)\sin(2\theta_2^*)\sin(\phi_1^*+\phi_2^*).
\end{align}
All kinematic variables entering these expressions are defined in
Sec.~\ref{sec:kinematic_conventions}. The numerical coefficients in
$f_{\rm ext}$, including the relative coefficient multiplying $A_E$ and the
scale $p_0=46~{\rm GeV}$, are selected by the search.

\section{Kinematic Conventions}
\label{sec:kinematic_conventions}

\paragraph{Associated production.}
For $e^+e^-\to Z(\to\ell^-\ell^+)H$, all angles are defined in the laboratory, equivalently center-of-mass, coordinate system before the decay lepton is boosted. Let $\hat p_{\rm beam}$ denote the chosen positive beam direction, $\hat p_{\rm beam}=(0,0,1)$ in the numerical evaluation, and let $\hat p_Z$ be the reconstructed $Z$ direction. The production angle is
\begin{align}
\cos\theta_Z=\hat p_{\rm beam}\cdot \hat p_Z .
\end{align}
The decay angles $(\theta^*,\phi^*)$ are defined for the negatively charged lepton after boosting it into the $Z$ rest frame. The $Z$ helicity-frame axes are
\begin{align}
\hat z=\hat p_Z,\qquad
\hat y=\frac{\hat p_{\rm beam}\times\hat z}
{|\hat p_{\rm beam}\times\hat z|},
\qquad
\hat x=\hat y\times\hat z,
\end{align}
so that $(\hat x,\hat y,\hat z)$ form a right-handed triad. If $\hat p_{\rm beam}\times\hat z$ is numerically degenerate, a fixed transverse fallback axis not parallel to $\hat z$ is used in the same construction. With $\vec p_{\ell^-}^{\,*}$ the negatively charged lepton momentum in the $Z$ rest frame,
\begin{align}
\cos\theta^*=
\frac{\vec p_{\ell^-}^{\,*}\cdot \hat z}{|\vec p_{\ell^-}^{\,*}|},
\qquad
\phi^*=\operatorname{atan2}
\left(\vec p_{\ell^-}^{\,*}\cdot\hat y,
\vec p_{\ell^-}^{\,*}\cdot\hat x\right),
\end{align}
with $\phi^*$ mapped to $[0,2\pi)$ in the numerical evaluation.
The charge-ordered laboratory-frame lepton asymmetries used by the extended
$ZH$ searches are
\begin{align}
A_E=\frac{E_{\ell^+}-E_{\ell^-}}{E_{\ell^+}+E_{\ell^-}},
\qquad
A_{p_T}=\frac{p_{T,\ell^+}-p_{T,\ell^-}}
{p_{T,\ell^+}+p_{T,\ell^-}},
\end{align}
together with the reconstructed $Z$ transverse momentum $p_T^Z$, rapidity
$y_Z$, and the lepton-pair separations $\Delta\phi_{\ell\ell}$ and
$\Delta\eta_{\ell\ell}$. Here $E_\ell$ and $p_{T,\ell}$ are measured in the laboratory frame, and the two-body kinematics imply
\begin{align}
A_E=-\beta_Z\cos\theta^*,\qquad
\beta_Z=\frac{|\vec p_Z|}{E_Z}.
\end{align}
The numerical scale in the selected expression is $p_0=46~{\rm GeV}$.

\paragraph{Four-lepton decay.}
For $pp\to H\to ZZ^*\to e^-e^+\mu^-\mu^+$, the flavor pairing is fixed:
\begin{align}
Z_1\equiv e^-e^+,\qquad Z_2\equiv \mu^-\mu^+ .
\end{align}
Let $\hat p_{\rm beam}$ now denote the positive proton beam direction, used only as a fixed transverse reference. In the reconstructed four-lepton rest frame, equivalently the rest frame of $p_{4\ell}=p_{Z_1}+p_{Z_2}$, define
\begin{align}
\hat z_i=\hat p_{Z_i}^{(4\ell)},\qquad
\hat y_i=\frac{\hat p_{\rm beam}\times\hat z_i}
{|\hat p_{\rm beam}\times\hat z_i|},
\qquad
\hat x_i=\hat y_i\times\hat z_i ,
\end{align}
Again a fixed non-collinear fallback axis is used in the degenerate case. The angles $(\theta_i^*,\phi_i^*)$ are defined for the negatively charged lepton from $Z_i$, boosted into the $Z_i$ rest frame:
\begin{align}
\cos\theta_i^*=
\frac{\vec p_{\ell_i^-}^{\,*}\cdot \hat z_i}
{|\vec p_{\ell_i^-}^{\,*}|},
\qquad
\phi_i^*=\operatorname{atan2}
\left(\vec p_{\ell_i^-}^{\,*}\cdot\hat y_i,
\vec p_{\ell_i^-}^{\,*}\cdot\hat x_i\right),
\end{align}
where $\ell_1^-=e^-$ and $\ell_2^-=\mu^-$. Since $\hat z_2=-\hat z_1$ in the reconstructed four-lepton rest frame, this convention gives $\hat x_2=\hat x_1$ and $\hat y_2=-\hat y_1$ away from the fallback case, so the combination $\phi_1^*+\phi_2^*$ is the signed dihedral angle between the two decay planes in this convention. The mass-ratio variable used in the main text is
\begin{align}
r_Z=\frac{m_{Z_2}}{m_{Z_1}+m_{Z_2}} .
\end{align}
Here $m_{Z_i}$ is the invariant mass of the fixed-flavor $Z_i$ candidate.

\section{Helicity Structure and Laboratory-Frame Mappings}
\label{sec:helicity_structure}

The main $ZH$ observable can be interpreted from the standard helicity
decomposition of $e^+e^-\to ZH$ followed by $Z\to\ell^+\ell^-$.  The rate can
be organized as
\begin{align}
d\sigma \propto \sum_{\lambda,\lambda'=0,\pm1}
\rho_{\lambda\lambda'}(\theta_Z)
D_{\lambda\lambda'}(\theta^*,\phi^*),
\end{align}
where the decay matrix contains phases
$D_{\lambda\lambda'}\propto e^{i(\lambda-\lambda')\phi^*}$.  The azimuthal
dependence is therefore restricted to harmonics with $|\lambda-\lambda'|=1,2$.
At linear order in the CP-odd deformation $\kappa$, the longitudinal--transverse
helicity interference contains the angular factor
\begin{align}
{\cal H}_{LT}
&\propto
\sin(2\theta_Z)\sin(2\theta^*)\sin\phi^*
\nonumber\\
&\propto
(\sin\theta_Z\cos\theta_Z)(\sin\theta^*\cos\theta^*)\sin\phi^* .
\end{align}
This factor is useful for interpreting the extended symbolic expression.  In
the same helicity language, the minimal symbolic observable recovers the
transverse--transverse CP-sensitive harmonic
\begin{align}
\sin^2\theta_Z\,\sin^2\theta^*\,\sin(2\phi^*),
\end{align}
while the extended expression adds a branch aligned with the signed
longitudinal--transverse factor.

The extended symbolic expression also contains laboratory-frame asymmetries.
In the $Z$ rest frame, neglecting lepton masses,
\begin{align}
E_\ell^*=\frac{m_Z}{2},\qquad |\vec p_\ell^{\,*}|=\frac{m_Z}{2}.
\end{align}
Boosting along the $Z$ direction gives
\begin{align}
E_{\ell^-}=\gamma_ZE^*(1+\beta_Z\cos\theta^*),\qquad
E_{\ell^+}=\gamma_ZE^*(1-\beta_Z\cos\theta^*),
\end{align}
and therefore
\begin{align}
A_E=-\beta_Z\cos\theta^* .
\end{align}
Thus $A_E$ is a monotonic proxy for the decay-side factor $\cos\theta^*$ in the
fixed-energy two-body setup.  In the evolved expressions it multiplies the
production-angle factor $\cos\theta_Z$,
\begin{align}
\sin\phi^*\sin\theta^*\cos\theta_Z A_E
\simeq
-\frac{\beta_Z}{2}\cos\theta_Z\sin(2\theta^*)\sin\phi^* .
\end{align}
This branch therefore provides the
$\cos\theta_Z\sin\theta^*\cos\theta^*\sin\phi^*$ dependence appearing in
${\cal H}_{LT}$.  At fixed $\sqrt{s}$, $p_T^Z=|\vec p_Z|\sin\theta_Z$, so the
smooth $p_T^Z$ prefactor supplies a production-angle weight.  With the nominal
30-bin Fisher estimator,
the isolated factors give
\begin{align}
\epsilon[\sin\phi^*\sin\theta^*A_E]&\simeq2.7\times10^{-4},\\
\epsilon[\sin\phi^*\sin\theta^*\cos\theta_Z]&\simeq3.6\times10^{-4},
\end{align}
whereas their product gives
\begin{align}
\epsilon[\sin\phi^*\sin\theta^*\cos\theta_ZA_E]
\simeq2.9\times10^{-2},
\end{align}
close to the efficiency of the explicit angular structure
$\cos\theta_Z\sin(2\theta^*)\sin\phi^*$.

The transverse-momentum asymmetry uses the same lepton charge ordering as the energy
asymmetry,
\begin{align}
A_{p_T}
=\frac{p_{T,\mu^+}-p_{T,\mu^-}}
{p_{T,\mu^+}+p_{T,\mu^-}} .
\end{align}
This directly measurable charge-ordered laboratory-frame lepton-$p_T$ asymmetry
is retained in the selected expression as part of the empirically optimized
laboratory-frame mapping, appearing within the same factor that multiplies the
CP-sensitive angular structure identified above. Together, the
helicity-interference angular factors, the $A_E$ identity, and the scan results
indicate that the evolved expression combines CP-sensitive angular harmonics,
charge-ordered lepton kinematic asymmetries, and a smooth $p_T^Z$ weight.

For the four-lepton channel, the same helicity-interference logic gives the
CP-sensitive signed decay-plane factor
\begin{align}
\sin(2\theta_1^*)\sin(2\theta_2^*)\sin(\phi_1^*+\phi_2^*) ,
\end{align}
with the angle convention of Sec.~\ref{sec:kinematic_conventions}.  This is the
angular kernel appearing in $O_{4\ell}$ in the main text; the role of the
additional mass-ratio envelope is tested by the ablations in Sec.~\ref{sec:search_robustness}.

\section{Symbolic Search and Robustness Checks}
\label{sec:search_robustness}

The symbolic search evolves compact analytic expressions of the available
event-level variables. In the LLM-guided evolutionary loop, the language model
is used only as a proposal engine for initial candidates, mutations, and
crossovers within the allowed function space. In the production runs reported
here, proposals were generated with the \texttt{qwen3.5-plus} model through an
OpenAI-compatible interface. The prompts specified the channel under study, the
available event variables, the allowed imports and elementary operations, and
the requirement that the output be a single executable analytic observable
function. No analytic matrix element or helicity amplitude formula was supplied
to the model.

All proposed expressions are then handled by a deterministic local evaluator:
they are parsed as Python functions, checked for disallowed imports, file
access, randomness, and global-state dependence, and evaluated on the fixed
score dataset. The language model does not assign the fitness, and its textual
reasoning is not used in the objective. Candidate expressions are evaluated by
the Fisher efficiency above, with small penalties for excessive expression size
and evaluation time.  The allowed operations are restricted to lightweight
analytic functions such as sums, products, protected ratios, powers, and
trigonometric combinations.  Candidates that return nonfinite values or
effectively constant outputs are rejected.

The objective used for selection is
\begin{align}
F[f]=\epsilon[f]-\lambda_{\rm comp}[f]-\lambda_{\rm run}[f],
\end{align}
where the penalties are subleading tie breakers.  For the production searches,
\begin{align}
\lambda_{\rm comp}[f]
&=\min\!\left(0.05,\frac{\max(0,N_{\rm AST}[f]-120)}{4000}\right),\\
\lambda_{\rm run}[f]
&=\min\!\left(0.02,\frac{t_{\rm eval}[f]/{\rm ms}}{200000}\right).
\end{align}
Here $N_{\rm AST}$ is the abstract-syntax-tree size of the candidate expression
and $t_{\rm eval}$ is the wall-clock evaluation time on the dataset.  The
search uses a population-based island strategy with elitism, fitness-weighted
parent selection, mutation, crossover, and periodic migration between
semi-independent subpopulations. In the default production configuration, the
search used eight islands with a total population of 160 candidate programs, an
80-generation budget, elitism of three candidates per generation, mutation and
crossover fractions of 0.75 and 0.25, respectively, and ring migration every
five generations with migration rate 0.10. Island and population decay were
enabled after generation 10 to concentrate the late-stage search on the best
surviving structures while retaining at least two active islands. These
mechanisms affect exploration, while the reported physics performance is always
the Fisher efficiency recomputed for the final expression.

For the $e^+e^-\to ZH$ study, a scan over the finite-difference step
\begin{align}
\delta =10^{-4},\;2\times10^{-4},\;5\times10^{-4},\;10^{-3},\;2\times10^{-3}
\end{align}
showed no visible instability in the reconstructed score.  Using
$\delta_{\rm ref}=5\times10^{-4}$ as reference, the score correlation was
numerically unity over the scan, the relative RMS shift of the raw score was
only $(3.4$--$5.1)\times10^{-8}$, and
\begin{align}
\frac{I_{\rm full}(\delta)}{I_{\rm full}(\delta_{\rm ref})}
=1+\mathcal{O}(10^{-9}).
\end{align}

A representative scan over the number of equal-population bins also preserved
the ordering between the evolved observable and the angular baselines.  Over
$N_{\rm bins}=10$--$60$, the best evolved observable varied only from
$0.0994$ to $0.1009$, while the strongest simple baseline $\sin(2\phi^*)$
varied from $0.0585$ to $0.0594$.  With five bins the corresponding values are
approximately $0.0953$ and $0.0572$, showing the expected coarser estimator but
unchanged ranking.

The dependence on feature representation was also checked in the $ZH$ channel.
With only
\begin{align}
x_{\rm min}=(\cos\theta_Z,\cos\theta^*,\phi^*)
\end{align}
the best evolved observable reaches $\epsilon\simeq0.076$ and already
contains the transverse-interference kernel.  Adding laboratory-frame
quantities such as lepton energy and transverse-momentum asymmetries,
$\Delta\phi_{\mu\mu}$, $\Delta\eta_{\mu\mu}$, $p_T^Z$, and $y_Z$ increases
the best efficiency to about $0.100$--$0.102$ without changing the core
$\sin(2\phi^*)$ structure.

The returned formula is not unique. Several high-performing $ZH$ candidates share the same qualitative structure,
\begin{align}
\left[
a\sin(2\phi^*)\sin^2\theta^*\sin\theta_Z
+\sin\phi^*\sin\theta^*\cos\theta_Z(bA_E+A_{p_T})
\right]W(p_T^Z),
\end{align}
while differing in numerical coefficients and smooth weighting functions. The same qualitative behavior is observed in the four-lepton channel: high-scoring candidates retain the CP-odd azimuthal kernel $\sin(\phi_1^*+\phi_2^*)$ and closely related polar-angle structures, while the smooth kinematic prefactor is less unique. In the selected compact representative the prefactor is the symmetric mass-ratio envelope $r_Z(1-r_Z)$. Direct ablations confirm this interpretation. With the nominal 30-bin estimator, $r_Z(1-r_Z)$ alone gives $\epsilon\simeq3.4\times10^{-5}$, while the angular kernel $\sin(2\theta_1^*)\sin(2\theta_2^*)\sin(\phi_1^*+\phi_2^*)$ gives $\epsilon\simeq1.9\times10^{-2}$, essentially the same as the full displayed observable. This supports the interpretation that the stable output of the search is the helicity-interference angular pattern rather than a unique algebraic representative or a uniquely informative mass envelope.

\section{Asymmetry Selection in the Four-Lepton Channel}
\label{sec:asymmetry_selection}

For the learned four-lepton observable $O_{4\ell}$, we tested a simple signed selection
\begin{align}
R_+:\; O_{4\ell}>O_{\rm cut},\qquad
R_-:\; O_{4\ell}<-O_{\rm cut},
\end{align}
with
\begin{align}
S_{\rm asym}=\sigma_{\rm int}(R_+)-\sigma_{\rm int}(R_-),\qquad
D_{\rm SM}=\sigma_{\rm SM}(R_+)+\sigma_{\rm SM}(R_-).
\end{align}
Here $\sigma_{\rm int}$ is obtained from $c_1$, while $\sigma_{\rm SM}$ is
obtained from $c_0$.
No continuum background is included in this signal-only diagnostic. A representative working point $|O_{4\ell}|>0.05$ gives $S_{\rm asym}/S_{{\rm asym},0}\simeq0.752$, $D_{\rm SM}/D_{{\rm SM},0}\simeq0.334$, a relative asymmetry-purity gain of about $2.25$, and a relative $S_{\rm asym}/\sqrt{D_{\rm SM}}$ gain of about $1.30$.
Scanning over $O_{\rm cut}$, the $S_{\rm asym}/\sqrt{D_{\rm SM}}$ measure
peaks near $O_{\rm cut}\simeq0.045$ with essentially the same maximal gain.
This scan is not part of the Fisher-information objective; it is included only
to show that the learned signed variable can also be used in a simple
interference-asymmetry selection.

\end{document}